\documentstyle[epsfig]{aipproc}

\begin{document}


\newcommand{\EEG}{\rm e^+ e^-\rightarrow \gamma\gamma}
\newcommand{\EEGG}{\rm e^+ e^-\rightarrow \gamma\gamma(\gamma)}
\newcommand{\EEGGG}{\rm e^+ e^-\rightarrow \gamma\gamma\gamma}
\newcommand{\EEEEG}{\rm e^+ e^-\rightarrow e^+e^-(\gamma)}
\newcommand{\LAMP}{ \Lambda_{+}}
\newcommand{\LAMM}{ \Lambda_{-}}
\newcommand{\LAMS}{ \Lambda_{6}}
\newcommand{\LAMPP}{ \Lambda_{++}}
\newcommand{\LAMMM}{ \Lambda_{--}}
\newcommand{\DSDW}{ \sigma(\theta)}
\newcommand{\MESTAR}{ m_{{\rm e^{\ast}}}}
\newcommand{\EELL}{\rm e^+ e^-\rightarrow l^{+}l^{-}}


\title{Limits on  sizes of fundamental particles 
and on gravitational mass of a scalar}

\author{ Irina Dymnikova$^*$, J\"urgen Ulbricht${^\dagger}$ and Jiawei Zhao$^{\ddagger}$
}
\address{$^*$Institute of Mathematics and Informatics, UWM
in Olsztyn, PL--10-561 Olsztyn, Poland\\
$^{\dagger}$Labor f\"ur H\"ochenergiephysik, ETH-H\"onggerberg, 
HPK--Geb\"aude, CH--8093 Z\"urich, Switzerland\\
$^{\ddagger}$Chinese University of Science and Technology, USTC, 
Anhui 230029 Hefei, P.R.China}

\maketitle

\begin{abstract}
We review 
the experimental limits on mass of excited fundamental particles
and contact interaction energy scale parameters $\Lambda$ for QCD, QED and  
electroweak reactions. In particular we have focused on the QED reaction    
$ \EEGG $ at the energies from 91GeV{} to 202GeV{}
using the differential cross-sections measured
by the L3 Collaboration from 1991 to 1999.
A global fit leads to lower
limits at $ 95 \% $ CL on 
$\Lambda > 1687$ GeV, which restricts the characteristic QED size of the
interaction region to $ R_{e} < 1.17 \times 10^{-17} $ cm. 
All the interaction regions are found to be smaller 
than the Compton wavelength 
of the fundamental particles. This constraint is used to estimate
a lower limit on the size of a fundamental particle related to gravitational
interaction, applying the model of self-gravitating particle-like structure
with the de Sitter vacuum core. It gives
$r_{\tau} \geq 2.3 \times {10^{-17}}$ cm and
$r_{e} \geq 1.5 \times 10^{-18} $ cm, if leptons get masses at the electroweak
scale, and $r_{\tau} \geq 3.3 \times {10^{-27}}$ cm,
$r_{e} \geq 4.9 \times 10^{-26} $ cm, as the most stringent limits required by causality arguments.
This sets also an 
upper limit on the 
gravitational mass of a
scalar $m_{scalar} \leq{ 154} $ GeV{} at the electroweak scale and 
\( m_{scalar} \leq \sqrt{3/8} m_{Pl} \) as the most stringent limit.
\end{abstract}      

\section*{Introduction}

In recent years,
three experimental approaches have been developed at LEP to
test the size of fundamental particles (FP).
First, a search is performed for excited
states of FP and a corresponding mass is estimated \cite{L3_LAT}.
Second, a characteristic scale
parameter $\Lambda$ is determined, constraining a
characteristic size of
the interaction region
for the reaction \cite{L3_LAT}.                       
Third, a form factor R \cite{form} has been
used to estimate the sizes of FP. 

None of the  measurements of the mass of excited FP, $\Lambda$ and R show 
a signal in frames of QCD, QED and electroweak theory, 
but it was possible to set stringent limits on all parameters.
In this paper we discuss, first, new experimental limits
from the L3 collaboration \cite{Budapest,L3Note2727}
in the QED part and prove that all the limits, set by QCD, 
QED and electroweak interaction on the characteristic sizes 
of the interaction region and form factors, are smaller 
than the Compton wavelengths 
$\lambdabar_c=\hbar/mc$ of FP. Second, we assume that 
whatever would be a mechanism of a mass generation,
a FP must have an internal core related to its                    
mass and a finite  
geometrical size defined by gravity.
To estimate it we apply de Sitter-Schwarzschild geometry 
which is the analytic globally regular modification of 
the Schwarzschild geometry corresponding to replacing
a singularity with a de Sitter vacuum core. 
This allows us to estimate an upper
 limit on a self-coupling $\lambda$ and on the gravitational mass of a scalar,
and to set the lower limits on sizes of FP as related
to gravitational interaction (\cite{us} and references therein).

\section*{EXPERIMENTAL LIMITS ON THE SIZES 
     OF FUNDAMENTAL PARTICLES }

\begin{figure} [b!] 
\center{
{\centering \begin{tabular}{cc}
\resizebox*{0.5\textwidth}{0.3\textheight}{\includegraphics{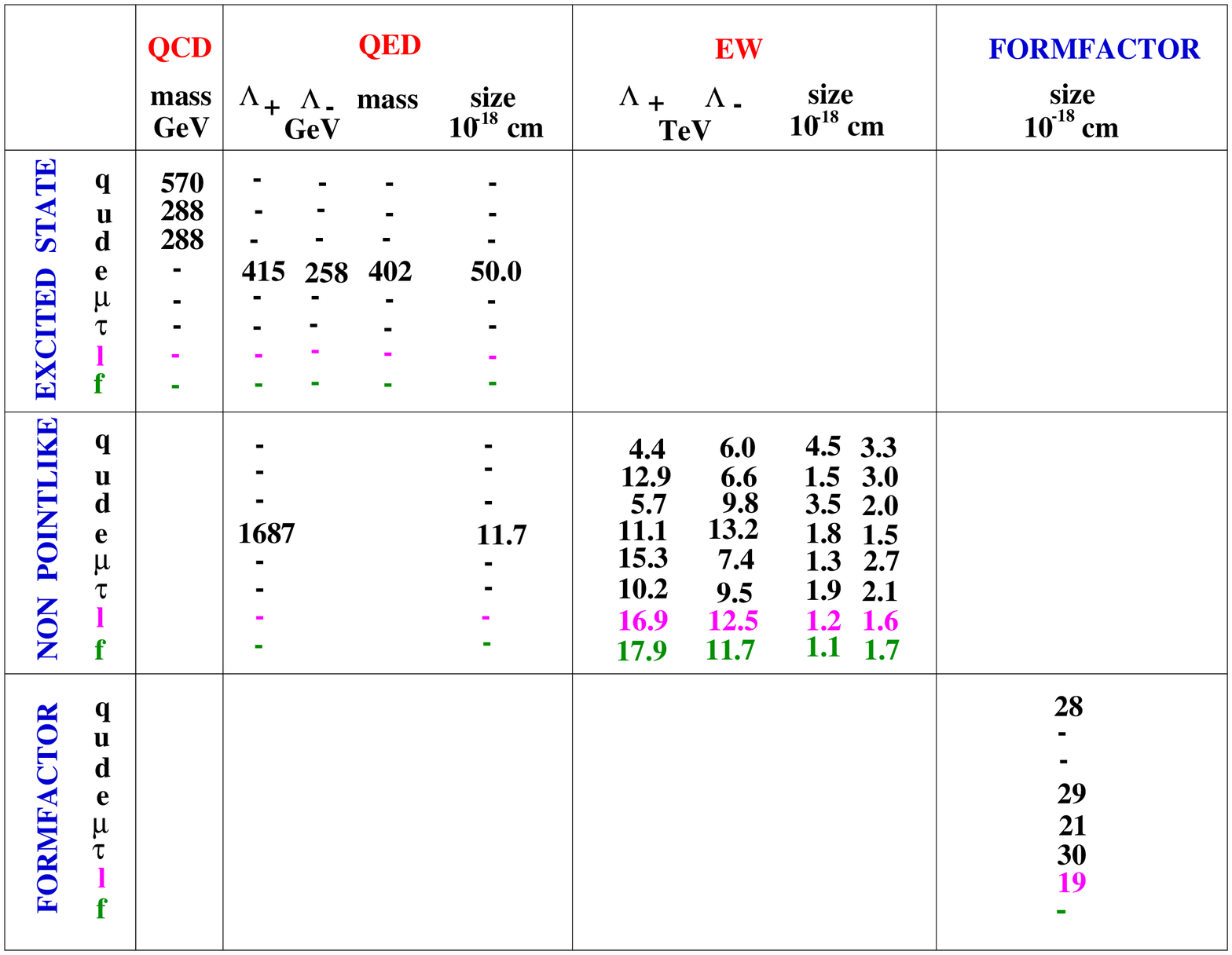}}&
\resizebox*{0.5\textwidth}{0.3\textheight}{\includegraphics{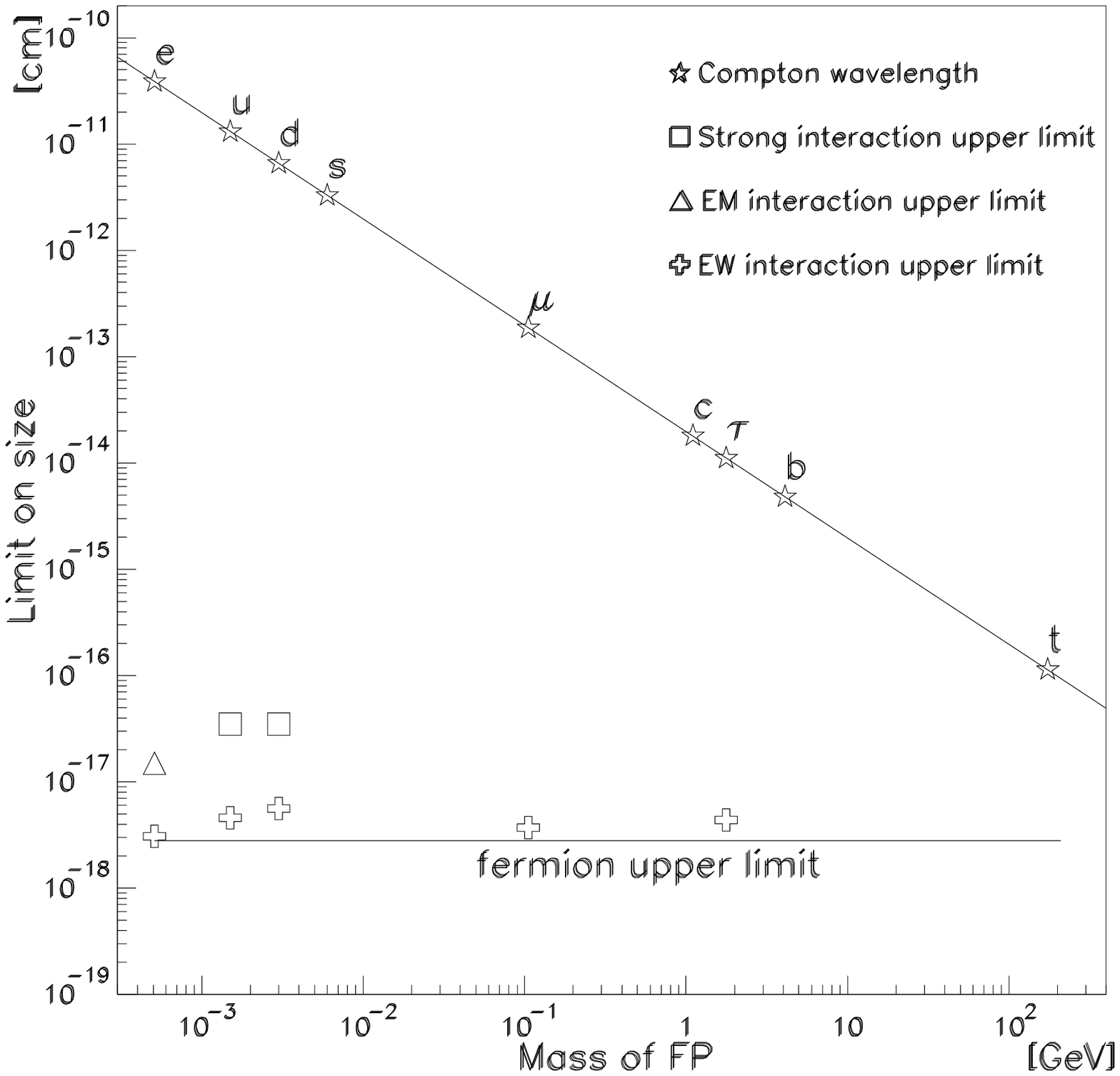}}\\
\end{tabular}\par}
}
\vspace{10pt}
\caption{In the left side the most stringent experimental limits of FP's are presented. The right side shows the comparison of Compton wavelengths of FPs with the current experimental limits measured for strong, electromagnetic and weak interaction.}
\label{summ}
\end{figure}

To test the finite size of fundamental particles,
experiments are performed to search for compositeness,
to investigate a non-pointlike
behavior or form factors R in strong, electromagnetic and electroweak
 interactions. Each interaction
is assumed to have   its characteristic energy scale                
related to the characteristic size of  interaction region. 
In the following sub-sections we review the experimental limits 
on excited particle masses, energy scales and form factors 
for all three interactions. 
We discuss the new QED results from the L3 experiment performed
during 1991 to 1999.
The most stringent limits measured from excited states
of fermions, non-pointlike couplings and form factors are shown in 
the left side of the Fig. \ref{summ}. In the right side they are compared with
the Compton wavelengths of FP.

{\bf Strong Interaction-}
To test  the color charge of the quarks,
the entrance channel
and the exit channels of the reaction
in the scattering experiment should
be dominated by the strong interaction. This condition
is fulfilled by the
 CDF $ p\bar{p} $
data \cite{CDF} which exclude excited quarks $ q^{*} $ with
a mass between $80$ and $570$~GeV at 95\%~CL. The
UA2 data \cite{UA2_EXQ} exclude $u^{*}$ and $d^{*}$ quark
masses smaller than $ 288 $~GeV at 90\%~CL.
In this case characteristic energy scale is given by
the mass of the excited quark. Associated characteristic size is
$r_q \sim       \hbar/(m_{q}^* c)<3.5\times 10^{-17}$ cm.

{\bf Electromagnetic Interaction-}
In the case of electromagnetic interaction
the process $ \EEGG $ is ideal to test the QED because
it is not interfered by the $ Z^{o} $ decay \cite{qed}. 
This reaction proceeds via the exchange of a virtual electron
in the t - and u - channels, while the s - channel is
forbidden due to angular momentum conservation.
Total and differential cross-sections for the process
$ \EEGG $,
are measured at the $\sqrt{s}$ energies from 91 GeV to 202 GeV
using the data \cite{L3_LAT,WU_THESIS,L3_2208,L3_125,L3_198}
and preliminary data \cite{Budapest,L3Note2727}
collected with the L3 detector from 1991 to
1999.

The agreement between the measured cross section
and the QED predictions is used to constrain
the existence of an excited
electron of mass $ m_{e^{*}} $ which
replaces the virtual
electron in the QED process \cite{LITKE}, or to constrain
a model with deviation from QED arising from an
effective interaction with non-standard
$ e^{+} e^{-} \gamma $ couplings and
$ e^{+} e^{-} \gamma \gamma $ contact terms \cite{EBOLI}.

An overall $\chi^2$ fit at 95\% CL \cite{L3Note2727} gives for the excited
electron~$\MESTAR >402$~GeV with
the QED cut-off parameters $ \LAMP > 415 $ GeV and
$ \LAMM > 258 $ GeV. $ \LAMP $ and $ \LAMM $
are mass cut-off parameters limiting the
existence of a heavy excited electron.
In the case of non-pointlike coupling
the cut-off
parameter $\Lambda$ limiting the scale
of the interaction is measured to be
$ \Lambda > 1687 $ GeV.
Characteristic size related to the case of interaction via
excited heavy electron is
$r_e\sim{\hbar / (m_{e^*}c) < } 5 \times 10^{-17} $ cm.
For the case of direct contact term interaction
$r_e \sim{ {(\hbar c )}/{\Lambda}}=1.17\times 10^{-17}$cm.
The behavior of
$ \chi^{2} $ as a function of $ \Lambda $ shows no minimum
indicating that the size of the interaction region
must be smaller than $r_e$.
                         
{\bf Electroweak Interaction-}
The $ ep $ accelerator HERA and the $ e^{+}e^{-} $
accelerator LEP  test excited and
non-pointlike couplings of quarks and leptons.
In the entrance channel the reaction proceeds 
via magnetic and weak interaction and in the exit channel
all three interactions participate. 

The H1 data \cite{H1_EXQ} give
for the $ q^{*}\rightarrow qg $ decay channel
a compositeness scale $ \Lambda $.
For a $ q^{*} $ of mass $ 100 $~GeV the
limit on $ \Lambda $ moves from $ 60 $ GeV
to $ 290 $ GeV.
In the $ q^{*}\rightarrow q + \gamma $ decay channel
the $ ep $ data exclude at 95 \% CL large regions of the
cross section times branching ratio
$\sigma(q^{*})\times BR(q^{*}\rightarrow q + \gamma) $
for $ q^{*} $ masses from $ 50 $ GeV to $ 250 $ GeV.
A similar search has been performed by the ZEUS
collaboration~\cite{ZEUS_EXQ}.

The H1 data \cite{H1_EXQ} describe for the
$ e^{*} $ case the electromagnetic and weak decay channel.
For the
$ \nu^{*} $ the channel $ \nu^{*} \rightarrow \nu \gamma  $
is measured. The experiment is able to set limits on the product
$ \sigma \times BR^{*} $ of the production cross section
and the branching ratio for different decay channels.
Big regions                               of the product
$ \sigma \times BR^{*} $ in the decay channels of
$ e^{*} \rightarrow e \gamma $,
$ e^{*} \rightarrow e Z^{0}  $,
$ e^{*} \rightarrow e W  $
$ \nu^{*} \rightarrow \nu \gamma  $ are excluded by the data in
the  $ e^{*} $ and $ \nu^{*} $ mass range up to $ 250 $ GeV
at $ 95 $\% CL. 

At the LEP, excited quarks and
leptons could be produced via
a $ Z^{0}, \gamma $ coupling to fermions.
The ALEPH investigated                                
the $q^*\rightarrow q+g$ and $q^*\rightarrow q+\gamma$
decay channels and 16 channels 
 from the
states $ l^{*}l $, $ \nu^{*} \nu $,
$ l^{*}\bar{l}^{*} $ and
$ \nu^{*} \bar{\nu^{*}} $ \cite{ALEPH_EXQ}.
No evidence for weak decay of excited quarks or
leptons has been found and stringent coupling
limits are set. By combining all radiative
channels under investigation
a  lower limit on the
compositeness maximal scale $ \Lambda > 16 $ TeV
for leptons could be established. L3 \cite{L3_2217} 
and OPAL \cite{OPAL_EXQ} report similar results. 

The search for non-pointlike coupling of the quarks and 
leptons is also performed with $ e^{+} e^{-} $
accelerators. The L3  searched at
center-of-mass energies between $ 130 $ GeV
and $ 172 $ GeV for new effects involving
four fermion vertices contact interactions
in all exit channels \cite{L3_FERMP}.
In particular
the L3 investigated the pure contact interaction
amplitudes $ e^{+} e^{-}\rightarrow f^{+} f^{-} $
and form factors for quark and leptons \cite{L3_FERMP,form}.

The most stringent limits for all three interactions
are summarized in the left side of the Fig.\ref{summ}.
 In the right side of the Fig. \ref{summ} a comparison is shown of the
 Compton wavelength $\lambdabar_c$ of FP 
with current experimental limits measured in
strong, electroweak and weak interaction. 
In conclusion all data show no signal 
\and Fig. \ref{summ} demonstrates that $\lambdabar_c$ 
is always bigger than the characteristic size of the interaction area. 
This experimental fact is used in the next section 
to estimate the sizes of FP and the mass of a Higgs scalar.

\section*{ CHARACTERISTIC SIZES  RELATED TO GRAVITY}

{\bf Selfgravitating particlelike structure with de Sitter vacuum core-}

De Sitter-Schwarzschild geometry originated from
replacing a black hole singularity with de Sitter
vacuum core.
The idea goes back to the 1965 paper by Gliner who interpreted the
Einstein cosmological term $\Lambda g_{\mu\nu}$ as a vacuum stress-energy
tensor $T^{vac}_{\mu\nu}=(8\pi G)^{-1}\Lambda g_{\mu\nu}=\rho_{vac}g_{\mu\nu}$ and suggested that it could be a final state 
in a gravitational collapse \cite{gliner}.
In the 80-s several solutions have been obtained by direct 
matching de Sitter metric inside to Schwarzschild metric 
outside of a junction surface \cite{numerical}.
All matched solutions have a jump in a metric at the junction 
surface, since the O'Brien-Synge junction condition $T^{\mu\nu}n_{\nu}=0$
is violated there \cite{werner}.
The exact analytic solution avoiding this problem and representing globally
regular de Sitter-Schwarzschild geometry, asymptotically Schwarzschild 
as $r\rightarrow\infty$ and asymptotically de Sitter as $r\rightarrow 0$,   
was found in the Ref.~\cite{IRINA1}.

The main steps to find this solution are to insert the spherically
symmetric metric
\begin{equation}
ds^{2}=e^{\nu }c^{2}dt^{2}-e^{\mu }dr^{2}-r^{2}
(d\theta^{2}+sin^{2}\theta d\phi ^{2})
\label{eq.7a}
\end{equation}
into the Einstein equations
\( R _{\mu \nu }-\frac{1}{2}Rg_{\mu \nu }=\frac{8\pi G}{c^{4}}T_{\mu \nu }
\)
which then take the form
\begin{equation}
\frac{-e^{-\mu }}{r^{2}}+\frac{\mu ^{\prime }
e^{-\mu}}{r}+\frac{1}{r^{2}}=\frac{8\pi G}{c^{4}}T_{t}^{t};\qquad
\label{eq.7b}
\frac{-e^{-\mu }}{r^{2}}-\frac{\nu ^{\prime }
e^{-\mu}}{r}+\frac{1}{r^{2}}=\frac{8\pi G}{c^{4}}T_{r}^{r}
\end{equation}

\begin{equation}
\frac{1}{2}e^{-\mu }(\nu ^{\prime \prime }+\frac{\nu ^{\prime 2}}{2}+\frac
{\nu ^{\prime }-\mu ^{\prime }}{r}-\frac{\nu ^{\prime }\mu ^
{\prime}}{2})=\frac{8
\pi G}{c^{4}}T_{\theta}^{\theta}=\frac{8\pi G}{c^{4}}T_{\phi}^{\phi}
\label{eq.7d}
\end{equation}
To match smoothly the de Sitter metric inside to the Schwarzschild metric
outside, the boundary conditions are imposed on the stress-energy
tensor such that \( T_{\mu \nu }\rightarrow 0 \) as \( r\rightarrow \infty \)
and \( T_{\mu \nu }\rightarrow \rho_{vac}g_{\mu\nu} \) as
 \( r\rightarrow 0  \), with \( \rho _{vac} \) as de Sitter vacuum density
at \( r = 0 \). For both de Sitter and Schwarzschild metrics the
condition \( \mu = -\nu \) is valid, which defines the class of spherically
symmetric solutions with the algebraic structure of the stress-energy tensor
\( T_{\mu \nu } \) such that
$T_{t}^{t}=T_{r}^{r}\,\, \textrm{and} \,\, 
T_{\theta }^{\theta }=T_{\phi }^{\phi }$.

The stress-energy tensor of this structure describes a spherically symmetric
vacuum,
invariant under the boosts in the radial direction (Lorentz rotations in $(r,t)$ plane)
\cite{IRINA1}, and can be interpreted as $r-$dependent cosmological term
\cite{lambda}.
It smoothly
 connects the de Sitter vacuum at the origin with the Minkowski
vacuum at infinity and generates the metric 
\label{alleqs9}
\begin{equation}
ds^2=\biggl(1-\frac{R_g(r)}{r}\biggr) dt^2 -
\biggl(1-\frac{R_g(r)}{r}\biggr)^{-1} dr^2 
- r^2( d{\theta}^2 + sin^2\theta d\phi^2)
\label{eq.9}
\end{equation}
where 
\label{alleqs10}
\begin{equation}
R_g(r)=\frac{2GM(r)}{c^2};~ ~ M(r)= \frac{4\pi}{c^2}\int_0^r{T^t_t(r)r^2 dr}
\label{eq.10}
\end{equation}
For any density profile, satisfying the conditions of needed
asymptotic behavior of a metric at the origin and of a finiteness
of a mass, this metric describes de Sitter-Schwarzschild geometry,
asymptotically Schwarzschild at infinity and asymptotically
de Sitter at the origin \cite{particle}.
In the model of Ref. \cite{IRINA1}
the density profile $T^t_t(r)= \rho(r) c^2$ has been chosen as
\label{alleqs12}
\begin{equation}
\rho = \rho_{vac} e^{ -4\pi\rho_{vac} r^3/3 m }=\rho_{vac}e^{-r^3/r_0^2 r_g}
\label{eq.12}
\end{equation}
which describes, in the semiclassical limit, vacuum polarization in the gravitational field
\cite{particle}.
Here
\label{alleqs14}
$
r_0^2 = {3 c^2}/{ 8 \pi G \rho_{vac}}$
is the de Sitter horizon,
$ r_g = 2 G m / c^2 $ is the Schwarzschild horizon, and \( m \) is
the gravitational mass of an object.

The metric  $g_{tt}= 1 - \frac{R_g(r)}{r}$ is shown
in Fig. 2. The fundamental difference from the Schwarzschild case
is that de Sitter-Schwarzschild black hole has two horizons, the black hole
horizon $r_{+}$ and the internal Cauchy horizon $r_{-}$ ($g_{tt}(r_{\pm})=0$).

\begin{figure} 
\center{
{\centering \begin{tabular}{cc}
\resizebox*{0.5\textwidth}{0.3
\textheight}{\includegraphics{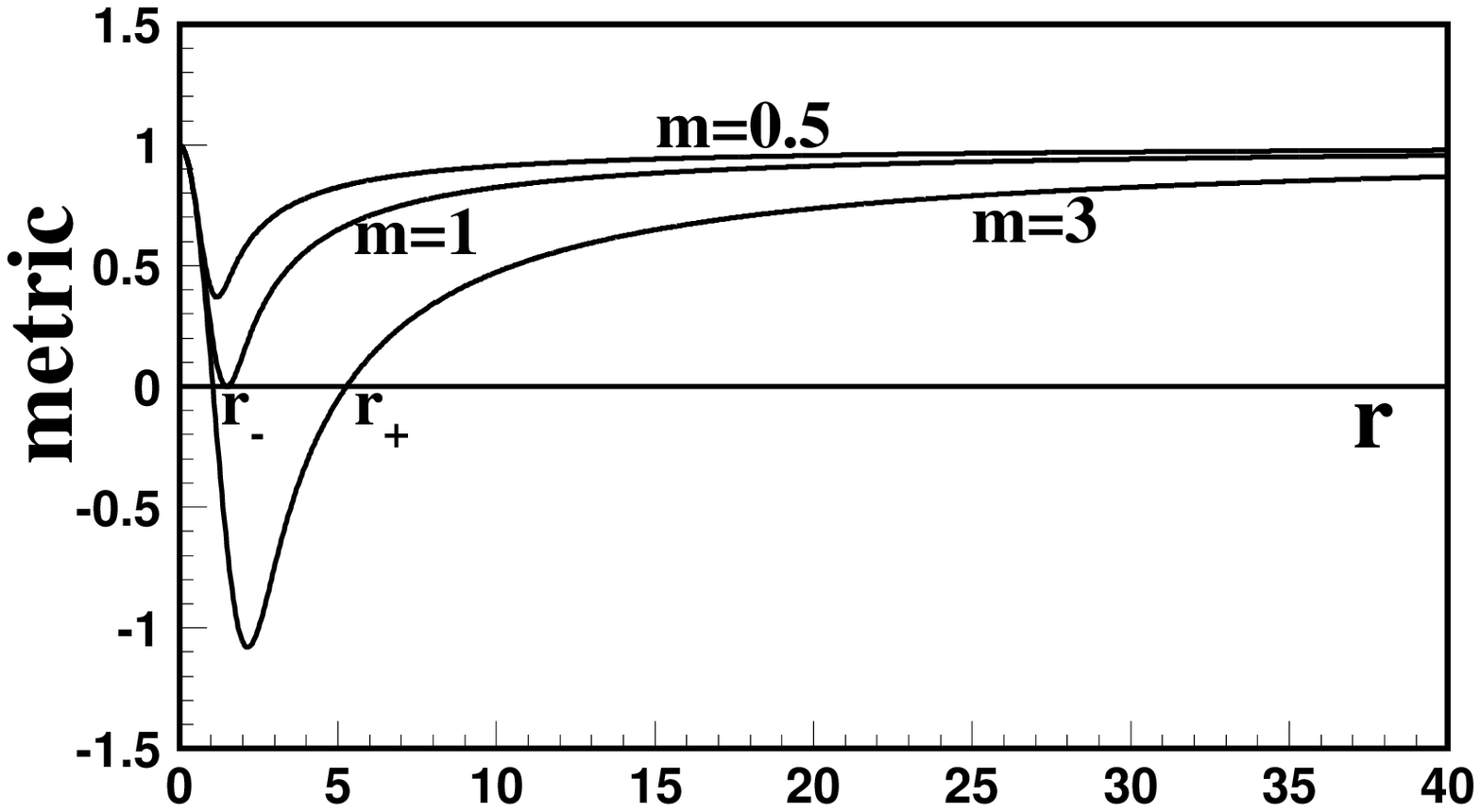}}&
\resizebox*{0.5\textwidth}{0.3\textheight}{\includegraphics{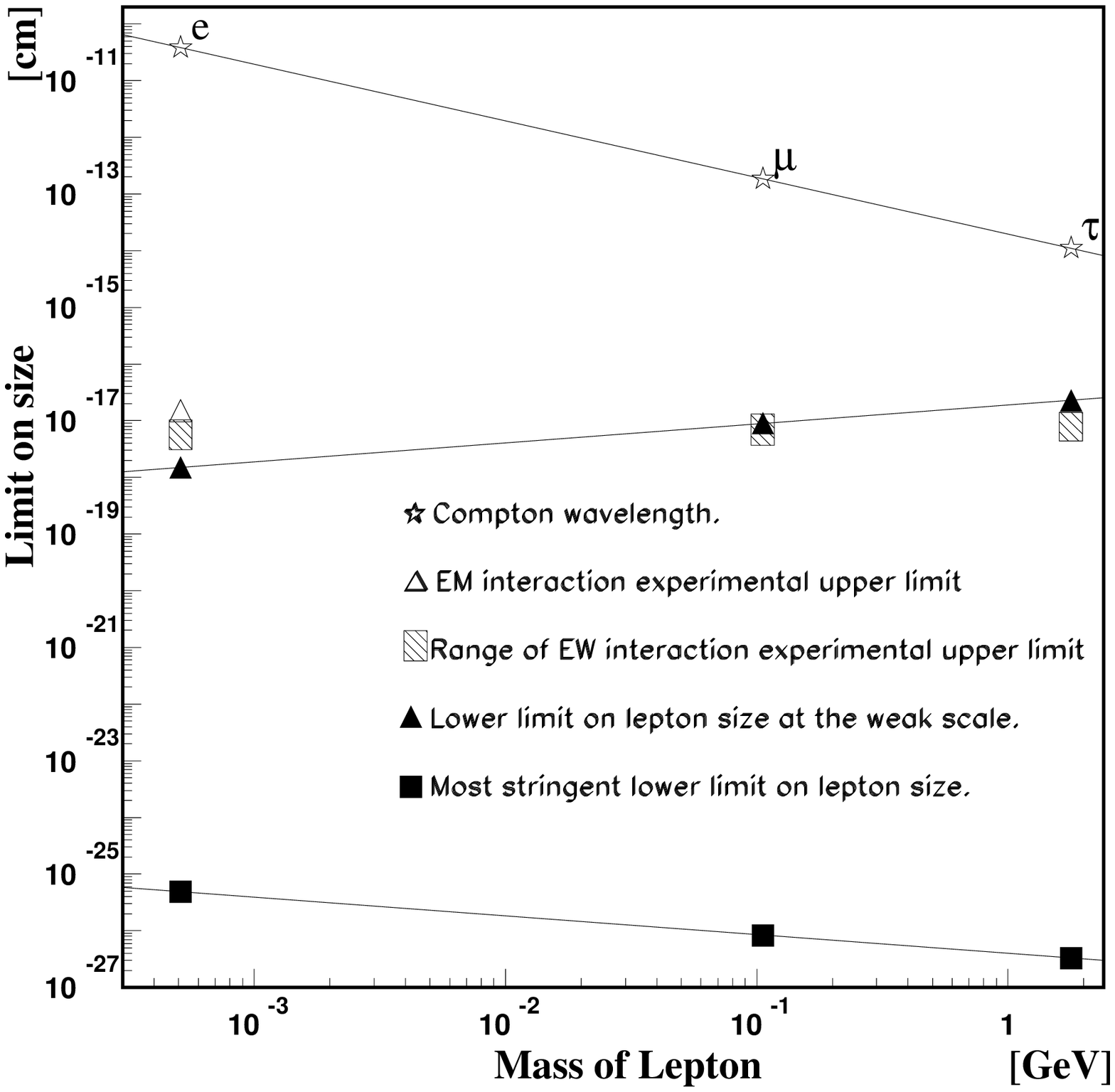}}\\
\end{tabular}\par}
}
\vspace{10pt}
\caption{ In the left side vacuum configurations described
by de Sitter-Schwarzschild geometry (\ref{eq.9})
$g_{tt}=1-R_g(r)/r$ are presented. The mass $m$ is normalized here
to $m_{cr}$. The case $m>1$ represents a black hole, $m=1$ its
extreme state, and $m<1$ a selfgravitating particlelike structure
with de Sitter vacuum core. The right side shows
the Compton wavelength of leptons as compared with experimental limits
of electromagnetic and weak interaction, and estimated lower
limits for the sizes of leptons.}
\label{fig.p3}
\end{figure}

The object is a black hole for
$m\geq m_{cr}\simeq{0.3 m_{Pl}\sqrt{\rho_{Pl}/\rho_{vac}}}$. It looses
its mass via Hawking
radiation until a critical mass \( m_{cr} \) is reached where
the Hawking temperature
drops to zero \cite{particle}.
At this point the horizons come together. The critical value
$m_{cr}$
puts the lower limit for a black hole mass.
Below $m_{cr}$ de Sitter-Schwarzschild geometry (\ref{eq.9})
describes a neutral
selfgravitating particlelike structure with 
$T_{\mu\nu}\rightarrow \rho_{vac}g_{\mu\nu}$
at the origin \cite{particle}. This
fact does not depend on particular form of a density 
profile \cite{particle}
which must only guarantee the boundary condition at the origin
 and the finiteness of the  mass as measured by a distant observer
\label{alleqs17}
\begin{equation}
m = 4\pi \int_0^{\infty} {\rho(r) r^2 dr}
\label{eq.17}
\end{equation}
De Sitter-Schwarzschild geometry has two characteristic surfaces
at the characteristic
scale $r\sim (r_{0}^2 r_{g})^{1/3}$ \cite{particle}. The first is the
surface of zero scalar curvature. The scalar curvature
\( R = 8 \pi GT \) changes its sign at the surface
\label{alleqs18}
\begin{equation}
r = r_s =
 \biggl ( \frac{m}{\pi \rho_{vac}}\biggr)^{1/3} =
 \frac{1}{\pi^{1/3}}   \biggl(\frac{m}{m_{Pl}}\biggr)^{1/3}
 \biggl(\frac{\rho_{Pl}}{\rho_{vac}}\biggr)^{1/3}l_{Pl}
\label{eq.18}
\end{equation}
which contains the most of the mass $m$.
Gravitational size  of a selfgravitating particlelike structure
can be defined by the radius $r_s$.
The second surface is related to
the strong energy condition of the
singularity theorems,
$  (T_{\mu\nu} - g_{\mu\nu}T/2)u^{\mu}u^{\nu}\geq 0$,
where $ u^{\nu} $ is any time-like vector. 
It is violated at the surface
of zero gravity
\label{alleqs19}
\begin{equation}
 r = r_c =
 \biggl ( \frac{m}{2\pi \rho_{vac}}\biggr)^{1/3} = \frac{1}
{(2\pi)^{1/3}} \biggl(\frac{m}{m_{Pl}}\biggr)^{1/3}
  \biggl(\frac{\rho_{Pl}}{\rho_{vac}}\biggr)^{1/3}l_{Pl}
\label{eq.19}
\end{equation}
Both these characteristic sizes represent modification of the Schwarzschild
gravitational radius $r_g$ to the case of a finite density $\rho_{vac}$
at the origin.

{\bf Sizes of lepton vacuum cores and an upper limit on a scalar mass-}

De Sitter-Schwarzschild particlelike structure cannot be applied
straightforwardly to approximate a structure of a FP
like an electron which is much more complicated. However, in the frame
of our assumption a mass of a FP is related to its gravitationally induced
core with de Sitter vacuum $\rho_{vac}$ at $r=0$.
This allows us to estimate the minimal geometrical size of a FP defined by
de Sitter-Schwarzschild geometry as a size of its vacuum core $r_c$, if we
know $\rho_{vac}$ and $m$.

In the context of spontaneous
symmetry breaking  the vacuum density $\rho_{vac}$ is related
to the vacuum expectation value $v$ of a Higgs field which gives
particle a mass $m=gv$, where $g$ is the relevant coupling to the scalar.
For a Higgs particle $g=\sqrt{2\lambda}$, 
where $\lambda$ is its self-coupling.
It is neutral and spinless, and we can approximate it by
de Sitter-Schwarzschild particlelike structure, identifying
$\rho_{vac}$ 
with the self-interaction of the Higgs scalar in the standard theory
\label{alleqs20a}
\begin{equation}
\rho _{vac}=\lambda v ^{4}/4
\label{eq.20a}
\end{equation}
We assume also that
the gravitational size of a particle \( r_s \) 
is restricted by its Compton
wavelength, \(r_s \leq \lambda \!\!\!\!-_c \).
This assumption 
is suggested by all experimental
data about limits on the sizes of FP ( See Fig.1 ).
This gives 
\label{alleqs21}
\begin{equation}
\frac{r_s}{\lambda \!\!\!\!-_c}=
\biggl(\frac{16\lambda}{\pi}\biggr)^{1/3} \leq 1
\,  \, \textrm{;} \, \, \lambda \leq \frac{\pi }{16}
\label{eq.21}
\end{equation}
With this limit on a self-coupling $\lambda$ we estimate an upper
limit on a Higgs scalar mass at the electroweak scale ($v=246$~ GeV) by
\label{alleqs22}
\begin{equation}
m_{scalar}\leq 154\, \textrm{GeV}
\label{eq.22}
\end{equation}
For a lepton getting its mass from the electroweak scale, 
its inner core is determined by this scale.
Putting Eq.(\ref{eq.20a}) into the Eq.(\ref{eq.19}) we get 
for a size of a vacuum core of a lepton with the mass $m_l$
\label{alleqs23}
\begin{equation}
{r_c}= \left(\frac{2m_l}{\pi\lambda v^4}\right)^{1/3}
\label{eq.23}
\end{equation}
Then the constraint on $\lambda$ (\ref{eq.21})
 sets the lower limits for the sizes of lepton vacuum cores by
$r_c^{(e)} > {1.5 \times {10^{-18}}}$~cm,
$r_c^{(\mu)} > {0.9 \times {10^{-17}}}$~cm, and
$r_c^{(\tau)} > {2.3 \times {10^{-17}}}$~cm.

To estimate the most stringent limit on $\rho_{vac}$ we take
 into account that quantum region
of localization \( \lambda \!\!\!\!- _{c} \) must fit within a casually
connected region confined by the de Sitter horizon \( r_0 \). The requirement
\( \lambda \!\!\!\!- _{c} \leq r_0 \) gives the limiting scale for a vacuum
density \( \rho _ {vac} \) related to a given mass \( m \)
\label{alleqs24}
\begin{equation}
\rho _ {vac} \leq \frac{3}{8 \pi} \left( \frac{m}{m_{Pl}}\right) ^{2}
\rho _ {Pl}
\label{eq.24}
\end{equation}
This condition connects a mass \( m \) with the scale 
for \( \rho_{vac} \) \
at which this mass could be generated in principle, whichever
would be a mechanism for its generation.

In this case 
we get from the Eq.(\ref{eq.19})
the most stringent, model-independent lower limit
for a size of vacuum core 

\label{alleqs28}
\begin{equation}
r_{c}>\left( \frac{4}{3}\right) ^{1/3}\left( \frac{m_{Pl}}{m_l}\right)
^{1/3}l_{Pl}
\label{eq.28}
\end{equation}

Inserting the masses of the leptons $m_l$ into Eq.(\ref{eq.28}), we find
$r_c^{(e)} >
 4.9 \times 10^{-26} $~cm, $r_c^{(\mu)} >
 8.3 \times 10^{-27} $~cm, and $r_c^{(\tau)} >
 3.3 \times 10^{-27} $~cm.

For a scalar we put Eq.(\ref{eq.20a}) and $m_{scalar}=\sqrt{2\lambda}v$ 
into Eq.(\ref{eq.24}) and get the limit on the vacuum expectation value  
$v\leq{\sqrt{3/\pi}m_{Pl}}$
valid for any self-coupling $\lambda$.
The restriction  
Eq.(\ref{eq.21}) gives an upper limit for a scalar mass 
$m_{scalar}\leq{\sqrt{3/8}m_{Pl}}$. These numbers give 
model-independent constraints 
for the case of particle production in the course of phase 
transitions in the very early universe. 

The limits on the sizes of FP are summarized in Fig.\ref{fig.p3}, 
compared to the $\lambdabar_c$ and to current experimental limits.
 The important is that the most stringent limits on sizes of FP
as estimated in the frame of de Sitter-Schwarzschild geometry, 
are much bigger than the Planck length $l_{Pl}\sim{10^{-33}}$ cm, what 
justifies our approach.
\vskip0.1in
{\bf ACKNOWLEDGMENTS}

We are  grateful to Samuel C. C. Ting for his strong support of this project,
and to Martin Pohl for stimulating discussions of this paper.

\end{document}